\documentclass[article]{agujournal2019}
\usepackage{url} %this package should fix any errors with URLs in refs.
\usepackage{lineno}
\usepackage[inline]{trackchanges} %for better track changes. finalnew option will compile document with changes incorporated.
\usepackage{soul}

\usepackage{bm}
\usepackage{bbm}
\usepackage{amssymb}
\usepackage{amsmath}

\draftfalse
\journalname{JGR: Planets}

\newcommand{\dt}[1]{\frac{\mathrm{d}#1}{\mathrm{d}t}}
\newcommand{\omg}{\bm{\omega}}
\newcommand{\omap}{\omega||\mathrm{MMOI}}

\begin{document}

\title{True Polar Wander on Dynamic Planets: Approximative Methods vs.~Full Solution}

% Example: \authors{A. B. Author\affil{1}\thanks{Current address, Antartica}, B. C. Author\affil{2,3}, and D. E.
% Author\affil{3,4}\thanks{Also funded by Monsanto.}}

\authors{Vojt\v{e}ch Pato\v{c}ka$^1$}

\affiliation{1}{Faculty of Mathematics and Physics, Department of Geophysics, Charles University, Prague, Czech Republic.
}

% Example: \correspondingauthor{First and Last Name}{email@address.edu}

\correspondingauthor{Vojt\v{e}ch Pato\v{c}ka}{patocka.vojtech@gmail.com}

%% Keypoints, final entry on title page.

%  List up to three key points (at least one is required)
%  Key Points summarize the main points and conclusions of the article
%  Each must be 100 characters or less with no special characters or punctuation and must be complete sentences

% Example:
% \begin{keypoints}
% \item	List up to three key points (at least one is required)
% \item	Key Points summarize the main points and conclusions of the article
% \item	Each must be 100 characters or less with no special characters or punctuation and must be complete sentences
% \end{keypoints}

\begin{keypoints}
\item True polar wander
\item Rotational bulge
\item nonlinear Liouville equation
\item numerical method
\end{keypoints}

%% ------------------------------------------------------------------------ %%
%
%  ABSTRACT and PLAIN LANGUAGE SUMMARY
%
% A good Abstract will begin with a short description of the problem
% being addressed, briefly describe the new data or analyses, then
% briefly states the main conclusion(s) and how they are supported and
% uncertainties.

% The Plain Language Summary should be written for a broad audience,
% including journalists and the science-interested public, that will not have 
% a background in your field.
%
% A Plain Language Summary is required in GRL, JGR: Planets, JGR: Biogeosciences,
% JGR: Oceans, G-Cubed, Reviews of Geophysics, and JAMES.
% see http://sharingscience.agu.org/creating-plain-language-summary/)
%
%% ------------------------------------------------------------------------ %%

\begin{abstract}
Almost three decades ago, the problem of long term polar wander on a dynamic planet was formulated and simplified within the framework of normal mode theory. The underlying simplifications have been debated ever since, recently in a series of papers by \citeA{Hu2017a,Hu2017b,Hu2019}, who clarify the role of neglecting short-term relaxation modes of the body. However, the authors still do not solve the governing equations in full, because they make approximations to the Liouville equation (LE). In this paper I use a time domain approach and for previously studied test loads I solve both the relaxation of the body and the LE in full. I also compute the energy balance of true polar wander (TPW) in order to analyze the existing LE approximations. For fast rotating bodies such as the Earth, I show that the rotation axis is always aligned with the maximum principal axis of inertia ($\omap$) once free oscillations are damped. The $\omap$ assumption is also re-derived theoretically. Contrary to previous beliefs, I demonstrate that it is not necessarily linked to the quasi-fluid simplification of the body's viscoelastic response to loading and rotation, but that it is an expression of neglecting the Coriolis and Euler forces in the equation of motion. For slowly rotating bodies such as Venus, the full LE together with energy analysis indicate that previous estimates of TPW in the normal direction need to be revisited. The numerical code LIOUSHELL is made freely available. 
\end{abstract}

\section*{Plain Language Summary}
The rotational poles of the Earth (the North and South poles) are slowly moving with respect to the Earth's surface. This is because large mass anomalies such as cold subducting slabs or big mountains tend to go away from the rotation axis and, conversely, large negative anomalies such as surface depressions tend to go towards the rotation axis. This phenomenon is referred to as true polar wander and it takes place also on other planets and moons. As a result of true polar wander, various prominent craters or mountains have different latitude now than they had at the time of their formation. In this paper I present a non-approximative method to compute true polar wander and for simple scenarios I compare the obtained polar motion to previously published solutions. The code used for the calculations is standalone, has a simple interface, and is made freely available. 

\section{Introduction}

Internal and external planetary processes such as mantle convection, deglaciation, or large meteorite impacts cause bodies to move as a whole relative to their rotation axis. The rotation poles of planets and moons thus wander on their surfaces, as shown by paleomagnetic, astrometric, and geodetic measurements \cite{Mitrovica2011}. Such reorientation in space is commonly referred to as true polar wander (TPW).

The TPW mechanism can be described as follows. Let us assume a rotating planet in hydrostatic equilibrium that is subject to a sudden loading, generating a positive geoid anomaly in effect. When viewed by an observer on the surface of the body, the rotation axis will start a circular motion around the new maximum principal direction of inertia (MMOI). As soon as these free oscillations dampen out, the new position of the rotation pole will be closer to the maximum principal axis of the load than in the initial state \cite<for free oscillations decay see>{Nakada2012}. Such motion of the rotation axis misaligns it with the initial rotational bulge, which opposes TPW in effect \cite<e.g.>{Matsuyama2014}. \citeA{Gold1955} argued that the stabilization is only transient because the rotational bulge will eventually adjust to the new direction of rotation. The body is then free to reorient further, and the process gradually proceeds until the load reaches the equator, or until it becomes fully isostatically compensated (i.e.~until its geoid signal disappears).

\citeA{Willemann1984} extended the model of \citeA{Gold1955} by incorporating a stabilization due to the so-called remnant (or fossil) rotational bulge. This way, the rotational bulge cannot adjust perfectly to the new rotation vector, $\omg$, because it is assumed that the rotational bulge partially formed when the body was hotter and its elastic lithosphere thinner, allowing thus for a larger flattening that is permanently imprinted into the present-day lithosphere. The remnant bulge acts as a static load and in the final state it counterbalances the driving load \cite<for a review, see>{Matsuyama2014}. 

Therefore, computing TPW requires an evaluation of the inertia tensor changes resulting from surface or internal loading, the subsequent isostatic compensation of the load, the viscoelastic relaxation of the rotational bulge, and also an estimate for the fossil constituent of the bulge. Until recently, the existing works that operated on geological timescales were based on the Love number theory \cite<e.g.>{Peltier1974}. Moreover, with the exception of \citeA{Nakada2007}, the published solutions were approximative in that they considered only the first-order Taylor expansion of the tidal Love number in the Laplace domain, i.e.~some of the body's viscoelastic relaxation modes were neglected. Such approximation is commonly referred to as the ``quasi-fluid'' approach \cite{Lefftz1991, Ricard1993}. Recently, \citeA{Hu2017b, Hu2019} analyzed the quasi-fluid approximation and by comparing the traditional solutions to their full-Maxwell approach they showed that the prediction of TPW speed on Earth and Mars were underestimated in previous studies.

The TPW itself is obtained from the conservation of momentum of a deformable body, known as the Liouville equation (LE). The LE is mutually coupled to the evolution of perturbations in inertia tensor. A number of works, in particular those related to Earth, have employed a linearized LE that is valid for small angle TPW \cite{Munk1960}. \citeA{Lefftz1991} showed that for excitation sources acting on a timescale much larger than the viscoelastic relaxation of the body, the LE reduces to a trivial formula, $\omg \times (\bm{I}\cdot\omg) = 0$ (or $\bm{I}\cdot\omg=\alpha \omg$ for the case with non-zero external torque, see \citeA{Ricard1993}, $\bm{I}$ is the inertia tensor, $\omg$ is the rotation vector). Hereafter I label this approximation as the $\omap$ approximation, but it is typically referred to as the ``quasi-fluid'' approximation as well, because its derivation was linked to the Taylor expansion discussed in the previous paragraph.

\citeA{Hu2017a, Hu2017b} develop a new approach, in which they iteratively solve the linearized LE to obtain a large angle TPW. \citeA{Hu2017b} argue that the validity of the $\omap$ approximation ``was discussed in detail in \citeA{Cambiotti2011} who showed that even for the Earth this assumption is not always appropriate''. However, such statement may be misleading, because \citeA{Cambiotti2011} actually employ the $\omap$ approximation, see their Eq.~7. They only show that the rotation axis is likely misaligned with the Maximum Inertia Direction of Mantle Convection (MID-MC), but they do assume that the rotation axis is aligned with the maximum direction of the total inertia tensor (MMOI). 

Ever since the problem was formulated in the Laplace domain, controversies arose about which relaxation modes can be safely neglected when predicting TPW of a layered viscoelastic body \cite<e.g.>{Vermeersen1996, Hu2017b}. Given the seemingly endless debate, and in view of the present-day CPU power, a question arises as to why one needs to make any approximations at all in the governing equations. One attempt in this direction is the paper by \citeA{Hu2017a}, but their model is not suitable for studies of bodies that contain layers with very different viscosities. The authors overcome this difficulty in their semi-analytical full-Maxwell model \cite{Hu2017b}, but the LE is still approximated in their work.

In this study, small viscoelastic deformations of a loaded rotating maxwellian body are computed by decomposing field variables into spherical harmonics. The governing equations are formulated directly in the time domain without omitting any relaxation modes. For hypothetical test loads, the full nonlinear LE is solved and compared to the $\omap$ approximation and also to the piecewise linear approximation of \citeA{Hu2017a}. 

The full LE solution is computationally reachable for up to millions of years on a single PC. For the TPW rate and colatitude evolution, both the $\omap$ and the piecewise linear approximations are shown to match the full LE solution. It is thus important to distinguish the two simplifications that often share the common label ``quasi-fluid'': i) simplification of the response of the body when the changes to the inertia tensor are evaluated, and ii) simplification of the LE. While the first was analyzed in detail by \citeA{Hu2017b} and proven to be a potential source of error, the latter is analyzed and rehabilitated here. 

To provide a rigorous framework for comparing the full LE solution with the approximative methods, I employ formulae from \citeA{Patocka2018} and compute the energy balance associated with the reorientation process. %An understanding of the energy balance is useful for finding inconsistencies, e.g.~when the employed principal moments of inertia do not correspond to the hydrostatic state of the underlying model, an issue raised by \citeA{Mitrovica2005} that is still debated, because the origin of the Earth's non-hydrostatic component is uncertain \cite{Hu2017b}.
A perhaps surprising result of the energetical analysis is that the $\omap$ approximation conserves energy. To support this finding, I analytically prove that the $\omap$ approximation is valid as long as Coriolis and Euler forces have a negligible effect on the body's deformation, i.e.~that its use can be justified independently of any simplifications related to the tidal Love number. For its simplicity and consistency, the $\omap$ equation can thus remain a convenient tool for modelling TPW on fast rotating planets.

TPW is often discussed in planetary science, because it results in changes of topography and gravity and can generate large tectonic stresses, i.e.~the primary observables of space missions. Several planets and moons were proposed to have reoriented in the past \cite<e.g.>{Nimmo2007,Schenk2008,Matsuyama2014}, and the topic is gaining increasing attention in recent years \cite<e.g.>{Keane2016,Tajeddine2017,Schenk2020}. Geomorphological evidence offers few clues about polar wander rate, and thus for planets other than Earth it is only the final reorientation that is typically investigated. 

Assessing the TPW rate, however, may be crucial for plausibility tests of the observation based reorientation scenarios. The TPW rate depends on the planet rheology, speed of rotation, shape, and on the amplitude and position of the driving load. In effect, the characteristic TPW timescale can vary by orders of magnitude, reaching or exceeding geological timescales on one hand, or being extremely short on the other. Its estimate should be included within geological interpretations of planetary reorientation, demanding a general and user friendly tool for TPW rate assessment. Recently, \citeA{Hu2019} provided a link between the TPW rate and the applicability of the fluid limit solution of \citeA{Matsuyama2007}, but estimates for reorientation timescales are still very scarce in the literature.

The code LIOUSHELL is a general tool for solving the LIOUville equation for viscoelastic SHells with an ELastic Lithosphere and can be used to predict TPW of both layered and smooth profiled bodies. It is based on a straightforward formulation in the time domain and the code is made freely available in Supplementary material. 

\section{Method}

Two sets of equations need to be solved in order to compute TPW for a given loading. The first set describes the viscoelastic response to surface or internal loads that drive TPW, as well as the formation of hydrostatic rotational bulge and its relaxation when the direction or speed of rotation changes. The second set is the LE, which takes deformation of the body as input and returns the changes in the rotation vector $\omg$ on output. The here employed numerical method was benchmarked and described in more detail in \citeA{Patocka2018}, and it is only briefly repeated below.

The small deformations of a hydrostatically prestressed viscoelastic spherical shell are obtained by integrating the following set of partial differential equations \cite<e.g.>{Tobie2008}:
\begin{eqnarray}
\nabla\cdot\bm{\tau} + \bm{f} = 0\, , \\ \label{pr}
\nabla\cdot\bm{u}=0\, , \\ \label{incomp}
\bm{\tau}^d-\mu(\nabla\bm{u}+(\nabla\bm{u})^\mathrm{T})=-\frac{\mu}{\eta}\int_0^t\bm{\tau}^d \mathrm{d}t' \, , \label{rheo}
\end{eqnarray}
where $\bm{\tau}$ is the Cauchy stress tensor, $\bm{\tau}^d$ is its deviatoric part, $\bm{f}$ is the body force, $\bm{u}$ is the displacement, $\mu$ is the elastic shear modulus, $\eta$ is the viscosity, and $t$ is the time. At the time $t{=}0$ the integral on the right-hand side of eq.~\eqref{rheo} is zero and Eqs \eqref{pr}--\eqref{rheo} give the initial elastic response.  

The body force $\bm{f}$ is given as the product of density and the negative gradient of the gravity potential. The gravity potential has three components: (i) reference gravitational potential $V_0$, (ii) centrifugal potential $\Psi(\omg,\bm{r})=0.5\left((\omg\cdot\bm{r})^2-|\omg|^2|\bm{r}|^2 \right)$ due to the rotation of the body, and (iii) perturbation $\Phi$ of the gravitational potential $V_0$ due to the deformation of the body. $V_0$ is the gravitational potential of a sphere of radius $a$ with a density profile $\rho_0(r)$, consisting of the solid spherical shell filled with a liquid core. 

As the initially spherical body deforms, the initial density $\rho_0(r)$ changes to $\rho(\bm{r},t)$, which can be expressed as a sum of density $\rho_0(r)$ and an Eulerian density increment $\delta\rho(\bm{r},t)$: 
\begin{equation}\label{drho}
\delta\rho(\bm{r},t) = \rho(\bm{r},t) - \rho_0(r) \cong -\bm{u}(\bm{r},t)\cdot\nabla\rho_0(r),
\end{equation}
where the displacement vector $\bm{u}$ describes the deformation of the body. The Eulerian density increment $\delta\rho$ is non-zero also at the outer surface and at any internal density interfaces, where it is equal to the density jump at each respective interface. Using $\delta\rho$, the perturbation potential $\Phi$ takes an integral form:
\begin{equation}\label{Phi}
\Phi(\bm{r},t)=-G\int_{v(t)} \frac{\delta\rho(\bm{r}',t)}{|\bm{r}-\bm{r}'|}\,\mathrm{d}v',
\end{equation}
where $v(t)$ is the volume occupied by the body at time $t$ (including the core). Neglecting the terms $O(|\bm{u}|^2)$ and $-\delta\rho\nabla(\Phi+\Psi)$, we obtain the following expression for the body force $\bm{f}$:
\begin{equation}\label{bforce}
\bm{f} = -\rho\nabla(V_0+\Phi+\Psi) \cong - (\bm{u}\cdot\nabla\rho_0)\bm{g}_0 - \rho_0\nabla\left(\Phi+\Psi\right) + \rho_0\bm{g}_0,
\end{equation}
where $\bm{g}_0=-\nabla V_0$. The laterally uniform and static contribution $\rho_0\bm{g}_0$ is counteracted by the hydrostatic prestress $p_0(r)$,
satisfying the equation $-\nabla p_0 + \rho_0 \bm{g}_0 = 0$.

The boundary condition at the outer surface is obtained from the force equilibrium condition, taking into account the pressure due to the deformation-induced topography and the presence of a surface load \cite<e.g.>{Soucek2016}:
\begin{equation}\label{bc1}
\bm{\tau}\cdot\bm{e}_r=\left(u_r[\rho_0]_a+\sigma^\mathrm{L}\right)\bm{g}_0,
\end{equation}
where $\bm{e}_r$ is the radial unit vector, $u_r$ is the radial component of displacement $\bm{u}$, $[\rho_0]_a = \rho_0(a)$ is the density jump at the surface, and $\sigma^\mathrm{L}$ is the surface mass density of the prescribed surface load. A similar condition can be imposed at the bottom boundary, on Earth corresponding to the core-mantle boundary (see \citeA{Soucek2016, Patocka2018}).

The temporal evolution of the angular velocity vector $\omg$ with respect to the body-fixed Tisserand frame \cite{Munk1960} is obtained from the LE with zero external torque,
\begin{equation}\label{Liou}
\frac{\mathrm{d}\omg}{\mathrm{d}t} = -\bm{I}^{-1} \cdot \left( \frac{\mathrm{d}\bm{I}}{\mathrm{d}t}\cdot\omg+\omg\times(\bm{I}\cdot\omg)\right),
\end{equation}
where $\bm{I}$ is the time dependent tensor of inertia. Eq.~\eqref{Liou} is the conservation of angular momentum $\bm{H} = \bm{I}\cdot\omg$, expressed in the body-fixed, i.e.~rotating geographic frame. Eqs \eqref{pr} -- \eqref{rheo} are coupled with eq.~\eqref{Liou} both through the displacement field $\bm{u}$, which is needed to compute the inertia tensor $\bm{I}$, and through the centrifugal potential $\Psi$, which depends on the angular velocity vector $\omg$. 

The inertia tensor $\bm{I}$ is computed via MacCullagh's formulae from the gravitational potential $\Phi$ and from the reference density profile $\rho_0$. Only the degree two components of spherical harmonics decomposition of $\Phi$ are needed to evaluate $\bm{I}$, which means that also eqs \eqref{pr}--\eqref{rheo} need to be solved for the degree two only, significantly reducing CPU time requirements in effect. Hereafter, solutions obtained from Eq.~\eqref{Liou} are labeled as ``full LE''.

\citeA{Lefftz1991} argued that when the variations of the driving force of TPW are much slower than viscoelastic relaxation of the body, the time derivative of the angular momentum $\bm{H}$ can be neglected in Eq.~\eqref{Liou}, resulting in the $\omap$ approximation:
\begin{equation}\label{qfLiou}
0 = \omg\times(\bm{I}\cdot\omg).
\end{equation}
In \ref{sec:App} I show an alternative derivation of Eq.~\eqref{Liou}, demonstrating that Eq.~\eqref{qfLiou} is equivalent to neglecting the Coriolis and Euler forces in Eq.~\eqref{pr}. This alone indicates that the range of applicability of the $\omap$ approximation may be broader than previously thought, because neglecting the Coriolis and Euler forces when evaluating the inertia tensor perturbations is more easily justified than neglecting the body's viscoelastic relaxation modes.

Since Eq.~\eqref{qfLiou} provides only the direction of $\omg$ and not it's amplitude, I use the conservation of momentum $\bm{H} = \bm{I}\cdot\omg = $ const to determine the amplitude, $\Omega := |\omg|$, at each time step.

Finally, I test also the piecewise linear approximation of \citeA{Hu2017a}, who iteratively solve an extended formulation of the traditional linearized LE,
\begin{eqnarray} \label{pwLiou}
\omega_1 / \Omega = \frac{I_{13}(t)}{C-A} + \frac{C}{\Omega (C-A)(C-B)}\dt{I_{23}} \nonumber\\
\omega_2 / \Omega = \frac{I_{23}(t)}{C-B} + \frac{C}{\Omega (C-A)(C-B)}\dt{I_{13}}  \\
\omega_3/ \Omega = -\frac{I_{33}}{C} + 1\, , \nonumber
\end{eqnarray}
where $A,B,C$ are the principal moments of inertia in the initial hydrostatic equilibrium. Since we do not assume tidal forcing here, the principal moments $A$ and $B$ are identical, $A=B$. \citeA{Hu2017a} iteratively transform system coordinates such as to keep the Z axis aligned with the $\omg$ vector in order to make Eq.~\eqref{pwLiou} applicable to large angle polar wander excursions, while the traditional linearized LE is valid only for small angles. Below, the solution obtained by Eq.~\eqref{pwLiou} together with the coordinate transformation is labelled as ``PWL''. In Section \ref{sec:Venus}, Eq.~\eqref{pwLiou} is replaced by Eq.~23 from \citeA{Hu2017b}.

\section{Model Setup}

I investigate two types of scenarios. In the first one the planet is assumed to spin fast and thus to have a large hydrostatic rotational bulge (e.g.~Earth, Mars). In the second scenario rotation is slow and the rotational bulge is small (e.g.~Venus). With the full LE, both scenarios can be treated in a unified manner. The approximative approaches for the two scenarios differ \cite{Hu2017b}.

To represent the first group, I employ the viscoelastic Earth model M3-L70-V01 \cite{Spada2011}, in which the viscosity profile is inferred from measurements of postglacial rebound. The second group is represented by the Venus model from \citeA{Spada1996,Hu2017b}. Before TPW is investigated, each model is rotated with a constant angular speed $\Omega_0$ until hydrostatic equilibrium is reached, and only then the body is loaded. Model parameters are summarized in Table \ref{tabModels}.

The code LIOUSHELL has already been tested against the benchmark example of \citeA{Spada2011}. In this study, however, I aim to provide a comparison of large angle TPW solutions obtained by the three variants of the LE, i.e.~Eqs \eqref{Liou}, \eqref{qfLiou}, and \eqref{pwLiou} respectively. To this end, two types of loading are considered: surface loads and fixed loads. After emplacing a surface load, the body deforms underneath and the load's geoid signal (and thus also its inertia tensor contribution) gradually weakens until it is negligible once isostasy is reached (and therefore TPW ceases). Fixed load, on the other hand, has a predefined inertia tensor contribution and it always ends up at the equator, because we do not consider a remnant rotational bulge here (it can be easily added as an additional fixed load).   

As the surface loading I use a spherical icecap centred at the colatitude 25° and longitude 75° \citeA<see Table 4 in>{Spada2011}, and I increase the ice density to $10^4$ kg/m$^3$ in order to obtain a large angle TPW. As the fixed loading I use the same icecap, but the load's surface traction is switched off, i.e.~$\sigma_\mathrm{L}$ is set to zero in Eq.~\eqref{bc1}, meaning that surface does not deflect below the ice cap (i.e.~no isostatic compensation is induced). Such a load is similar to the point loads studied in \citeA{Hu2017a} and \citeA{Hu2017b}, but it can also represent e.g.~mantle convection or other internal sources that can be schematized by prescribing a given inertia tensor contribution \cite<see e.g.>{Cambiotti2011}.

Note that only the degree 2 components of the ice cap loading are used in the benchmark of \citeA{Spada2011} and for consistency the degree 0 is neglected also here. Therefore, the total mass of the cap mass is zero, because only the degree 0 spherical harmonic component of any surface loading contributes to its total mass (in this case, the degree 0 contribution would make $3.9\times10^{19}$ kg). Accounting for the load mass would not change the below results significantly, with the exception of the obtained change in the length of the day at the instant of loading. Assuming a zero mass cap can also be understood as forming the load self-consistently from the Earth material, e.g.~by freezing water that is already on the planet for the case of ice caps. Note also that in this study I do not employ any correction to the initial tensor of inertia, while to fit the original benchmark a correction is needed \cite<see>[for a detailed discussion]{Patocka2018}.

To represent a slowly rotating body I study the mega wobble on Venus, similarly to \citeA{Spada1996,Hu2017b}. Model parameters are listed in Table \ref{tabModels}. After reaching hydrostatic equilibrium for a constant rotation, the body is loaded with a point mass of $-5\times 10^{18}$ kg \cite{Spada1996,Hu2017b}. Load parameters are listed in Table \ref{tabLoad}.

\begin{table}
\caption{Model parameters}
\centering
\begin{tabular}{l c c c}
Earth model & $\Omega_0=7.2921{\times}10^{-5}$ s$^{-1}$ & & \\
\hline
Outer radius (km)  & Density (kg/m$^3$) & Shear modulus (Pa) & Viscosity (Pa s)  \\
\hline
   6371  & 3037 & $0.50605\times10^{11}$ & $\infty$ \\
   6301  & 3438 & $0.70363\times10^{11}$ & $1\times 10^{21}$ \\
   5951  & 3871 & $1.05490\times10^{11}$ & $1\times 10^{21}$ \\
   5701  & 4978 & $2.28340\times10^{11}$ & $2\times 10^{21}$\\
   3480  & 10750 & $0$ & $0$ \\
\hline
Venus model & $\Omega_0 = 3.0009{\times}10^{-7}$ s$^{-1}$ & & \\
\hline
Outer radius (km)  & Density (kg/m$^3$) & Shear modulus (Pa) & Viscosity (Pa s)  \\
\hline
   6052  & 2900 & $0.36\times10^{11}$ & $\infty$ \\
   6002  & 3350 & $0.68\times10^{11}$ & $0.6\times 10^{21}$ \\
   5500  & 3725 & $0.93\times10^{11}$ & $1.6\times 10^{21}$ \\
   5200  & 4900 & $2.07\times10^{11}$ & $6.4\times 10^{21}$\\
   3250  & 10560 & $0$ & $0$ \\
\hline
\end{tabular}\label{tabModels}
\end{table}

\begin{table} 
\caption{Load parameters}
\centering
\begin{tabular}{l c c c c c}
\hline
Label  & Mass (kg) & Colatitude (°) & Longitude (°) & Isostasy & Cap density$^{b}$ (kg/m$^3$) \\
\hline
   ES  & 0 $^{a}$ & 25 & 75 & yes & 10$^4$ \\
   EF  & 0 $^{a}$ & 25 & 75 & no & 10$^4$ \\
   VP  & $-5\times 10^{18}$ & 45 & 0 & no & \\
\hline
\end{tabular}
{$^{a}$The cap is 1500 m tall and has half-extension of 10°, see Table 4 in \citeA{Spada2011}. Only its degree two components are considered and its total mass is thus zero (see text for discussion). $^{b}$Note that in the benchmark of \citeA{Spada2011} the ice density is 931 kg/m$^3$ and here it is artificially increased in order to induce large angle TPW.}
\label{tabLoad}
\end{table}

\section{Results: Fast Rotating Planets}

\begin{figure}
    \centering
    \includegraphics[width=1.0\textwidth]{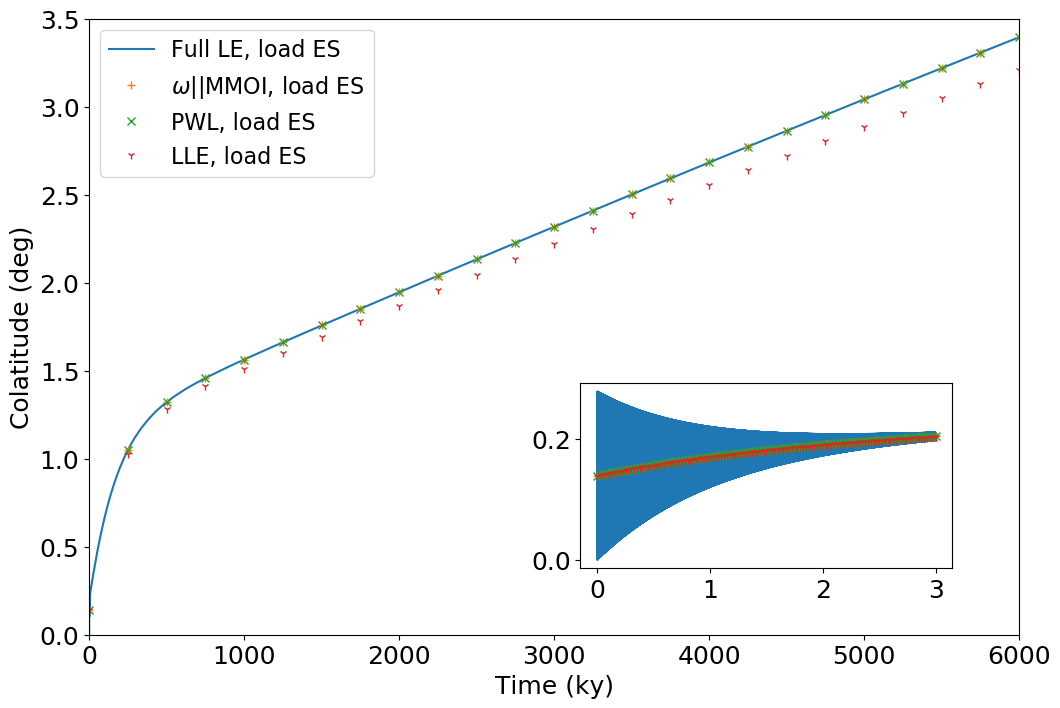}
    \caption{The evolution of colatitude for Earth loaded by the spherical cap ES (see Table \ref{tabLoad}). The approximative solutions ($\omap$ and PWL) both converge to the full LE solution once the free oscillations are damped. The figure inset displays the first few ky of the evolution during which the free oscillations rapidly decay. For illustration purposes, also the traditional linearized LE solution is plotted (red crosses).}
    \label{figColES}
\end{figure}

Hydrostatic Earth is instantaneously loaded at the time $t{=}0$ with the spherical cap ES (Table \ref{tabLoad}). In Fig.~\ref{figColES} the evolution of colatitude of the Earth's rotation axis in the body-fixed frame is plotted. Emplacing the load first induces free oscillations, which are damped within a few ky (see the blue line in the inset of Fig.~\ref{figColES}). The TPW rate is initially relatively rapid, but it soon decays as isostatic accommodation of the load proceeds.

The $\omap$ and the piecewise linear (PWL) approximations are mutually indistinguishable and both converge to the full LE solution once the free oscillations dampen out. Note that the colatitude of $\omg$ is non-zero at the time $t{=}0$ for the approximative solutions. This is because the vector $\omg$ is iteratively adjusted to instantaneously match the new position of MMOI (cf.~Eq.~\eqref{qfLiou}). For illustration purposes, I plot also the solution of the traditional linearized LE (red crosses). As expected, the LE is inaccurate for the large angle TPW. For the piecewise linear approach of \citeA{Hu2017a} this deficiency is overcome by the iterative transformation of the coordinates via matrix $\bm{Q}$ in their Eq.~15. 

\begin{figure}
    \centering
    \includegraphics[width=1.0\textwidth]{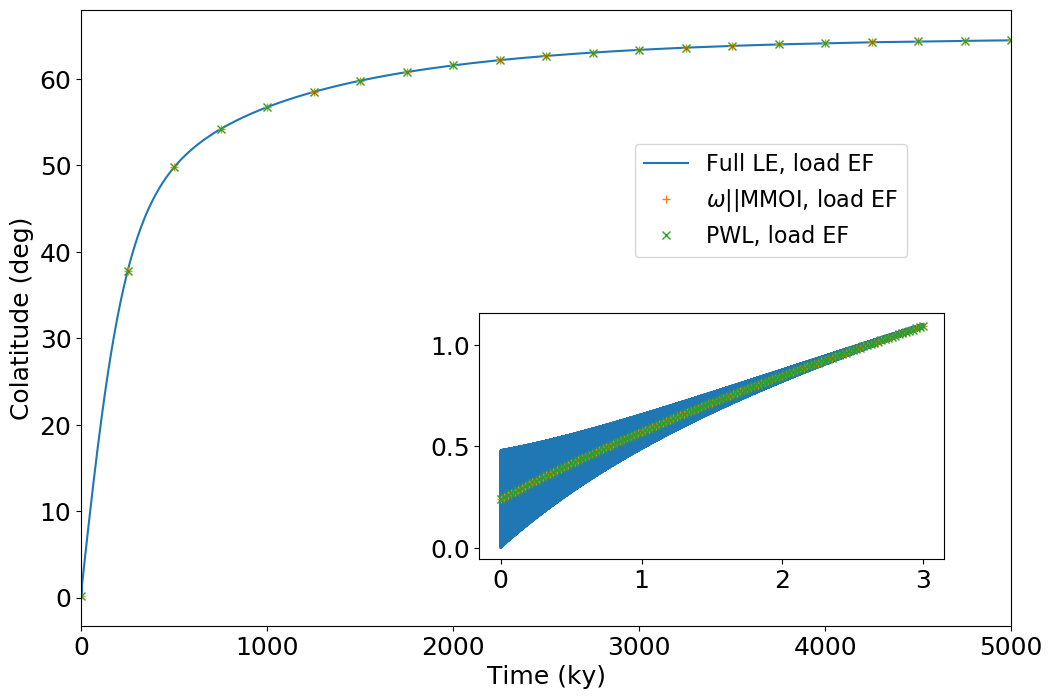}
    \caption{Same as Fig.~\ref{figColES}, only here the effect of isostasy is removed. Without sub-load deformation, the planet quickly reorients to its final state in which the load is at the equator, i.e.~in which $\omg$ reaches the 65° colatitude. Again, both the approximative solutions converge to the full LE solution.}
    \label{figColEF}
\end{figure}

In Fig.~\ref{figColEF} the above experiment is repeated, only this time the cap is artificially attached to the body, not inducing any isostatic sub-load relaxation (Table \ref{tabLoad}). Comparison of the TPW amplitude in Figs \ref{figColES} and \ref{figColEF} illustrates the importance of isostasy in reducing the load's contribution to geoid.

Similarly to Fig.~\ref{figColES}, both the $\omap$ and PWL approximations converge to the full LE solution as soon as the free oscillations dampen out, indicating that such a convergence is of a general nature for bodies with large bulge to load ratio. After a few My, the load reaches its final position at the equator and reorientation is completed.

\begin{figure}
    \centering
    \includegraphics[width=1.0\textwidth]{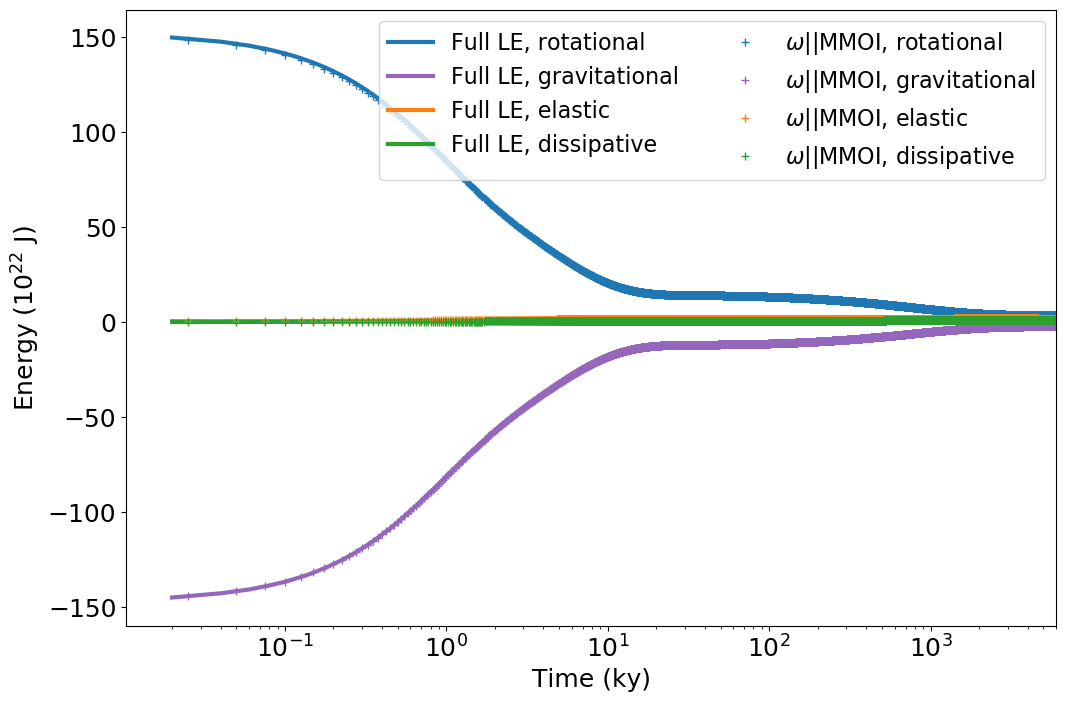}
    \caption{Evolution of the rotational, gravitational, elastic, and dissipative energies for the TPW process studied in Fig.~\ref{figColES}, i.e.~for the model Earth loaded with the spherical cap ES (Tables \ref{tabModels} and \ref{tabLoad}).}
    \label{figENESfull}
\end{figure}

\begin{figure}
    \centering
    \includegraphics[width=1.0\textwidth]{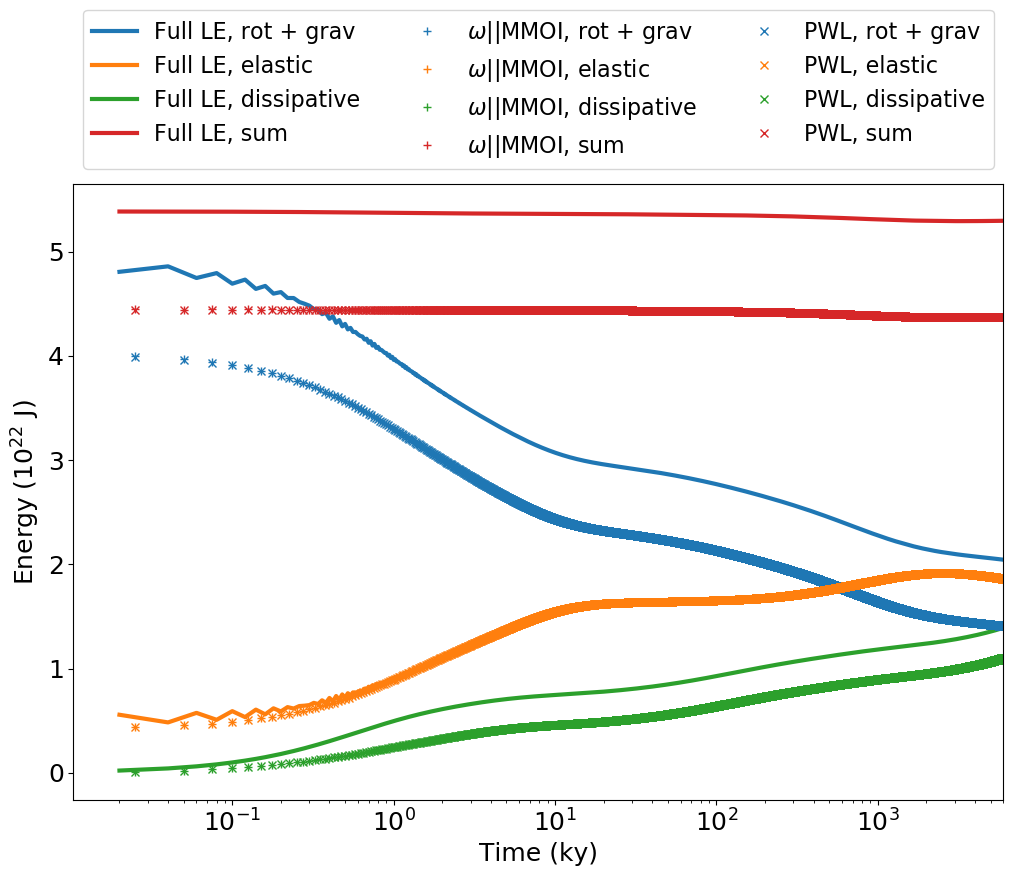}
    \caption{Same as Fig.~\ref{figENESfull}, only here the rotational and gravitational energy are combined in order to see the evolution of the remaining energy constituents in more detail.}
    \label{figENES}
\end{figure}

In Fig.~\ref{figENESfull} I show the evolution of the gravitational, rotational, elastic, and dissipative energy associated with the TPW process studied above. Earth's polar wander is dominated by the decay of the rotational energy, which is mostly counterbalanced by the increase of the gravitational energy (blue and purple lines \ref{figENESfull}). To have a better insight into the behaviour of the other energy constituents, it is thus convenient to sum the rotational and gravitational energy into a single contribution (Fig.~\ref{figENES}).

In Figs \ref{figENESfull} and \ref{figENES} the energy is plotted with respect to the initial hydrostatic state. The evolution of the individual energy components is similar for the full LE and for the $\omap$ (and PWL) approximation. There are only two notable differences: i) At the time $t{=}0$ the increase in the rotational and gravitational energy with respect to the hydrostatic state differs. This is because for the full LE I enforce the conservation of momentum upon load emplacement by changing the amplitude of $\omg$,  $\omega_3(t{=}0) = \Omega_0 (\bm{I}(t{=}0)\cdot\bm{\Omega}_0) / (\bm{I}_\mathrm{h}\cdot\bm{\Omega}_0)$, while for the $\omap$ approximation Eq.~\eqref{qfLiou} is applied at the time $t{=}0$, which slightly changes also the direction of $\omg$. ii) Dissipation is larger for the full LE solution, because dampening of the free oscillations contributes to it. With respect to the changes in the rotational and gravitational energy, however, this dissipative component is very small (cf.~Fig.~\ref{figENESfull}).

The energy is conserved to within less than 2\% error for the plotted solutions, which is less than 0.1\% error relative to the largest components of the budget (cf.~Fig.~\ref{figENESfull}). Note that the $\omap$ approximation also conserves energy, because the Euler and Coriolis forces are consistently neglected in both the underlying governing Eqs \eqref{pr} and \eqref{qfLiou}. The approximation does, nevertheless, completely omit the wobbling process and the dissipation released through its damping, which may be a major component of the energy balance in some scenarios (see Section \ref{sec:Venus}). 

Note that the interplay of the individual components in Fig.~\ref{figENES} is quite complex. This may help to identify inconsistencies such as using a model that is initially not in hydrostatic equilibrium. When that happens, the TPW process becomes dominated by the continuing relaxation of the rotational bulge, diminishing the role of elastic energy in the subsequent TPW process. 

\begin{figure}
    \centering
    \includegraphics[width=1.0\textwidth]{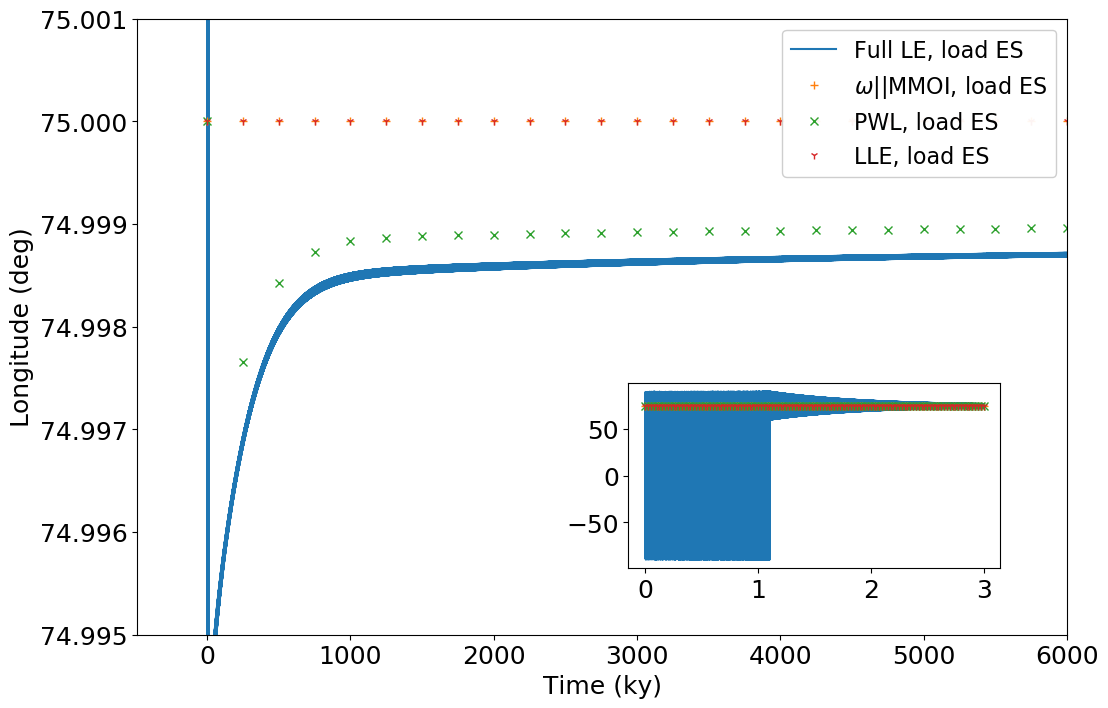}
    \caption{The evolution of longitude of $\omg$ for the model Earth loaded with the spherical cap ES (Table \ref{tabLoad}). During the first ca.~1 ky the body-fixed frame's $z$-axis lies within the circular trajectory of the free wobble. For figure clarity I define the longitude as $\arctan(\omega_2/\omega_1)$, i.e.~its value spans the range of -90° to 90° within that the first 1 ky instead of -180°to 180°. }
    \label{figLongitES}
\end{figure}

While the colatitude and energy evolution is indistinguishable for the $\omap$ and PWL approximations, the evolution of longitude slightly differs. As illustrated in Fig.~\ref{figLongitES}, the PWL approximation is better at capturing the offset of longitude that is caused by the initially rapid TPW.

\section{Results: Mega Wobble on Slowly Rotating Planets}\label{sec:Venus}

Next I analyze a planet with slow rotation. In this scenario, the stabilization effect of rotational bulge is significantly reduced due to its small size (less than a meter difference in the polar vs equatorial radius for the model of Venus). Free oscillations then have a much larger period and can no longer be neglected, because a relatively small load may trigger a large amplitude wobbling \cite{Spada1996}.

\begin{figure}
    \centering
    \includegraphics[width=1.0\textwidth]{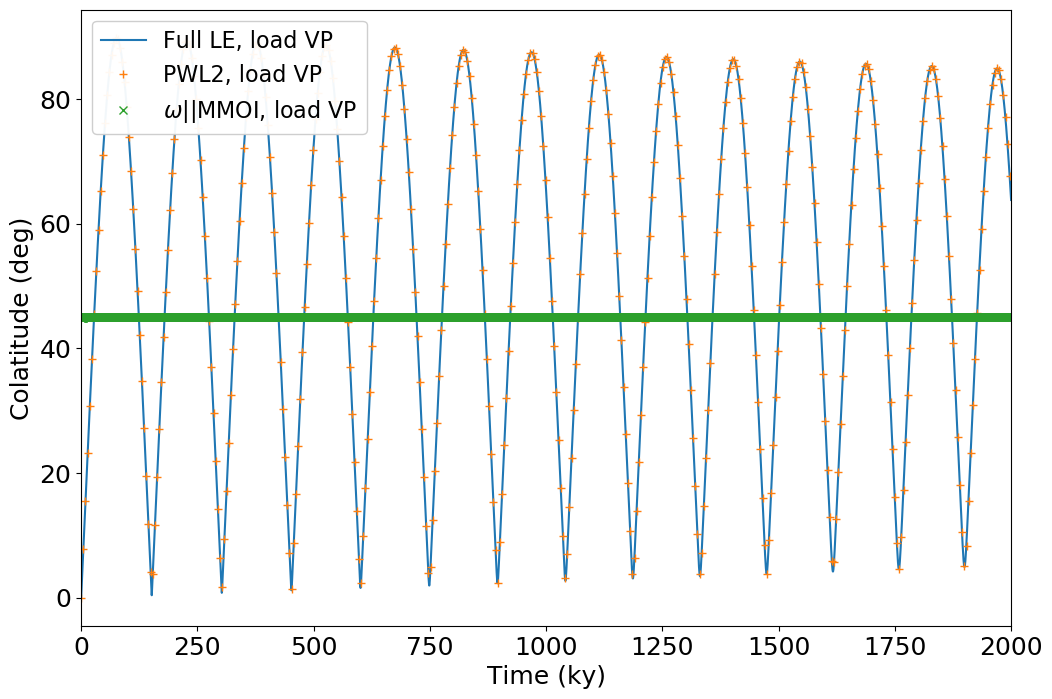}
    \caption{The evolution of colatitude for model Venus, loaded by the negative point mass anomaly VP (Table \ref{tabLoad}).}
    \label{figColVP}
\end{figure}

For slowly rotating bodies, Eq.~\eqref{pwLiou} is inappropriate and \cite{Hu2017b} derive a different approximation in their Eq.~23. In Fig.~\ref{figColVP} I show the evolution of colatitude after loading the hydrostatic Venus with the point mass VP (Tables \ref{tabModels} and \ref{tabLoad}). Again, I compare the full LE solution with the $\omap$ approximation and with the solution obtained via Eq.~23 in \citeA{Hu2017b} (labelled as PWL2, see the green symbols in Fig.~\ref{figColVP}).

Clearly, the $\omap$ approximation is not suitable for modelling the mega wobble. In fact, wobbling is by default omitted within its governing Eq.~\eqref{qfLiou}. The $\omap$ approximation merely captures the final position of the rotation vector after wobbling is damped, with the final position being reached nearly instantaneously, because the load amplitude ($5{\times}10^{18}$ kg) is significantly larger than the mass of the rotational bulge.

\begin{figure}
    \centering
    \includegraphics[width=1.0\textwidth]{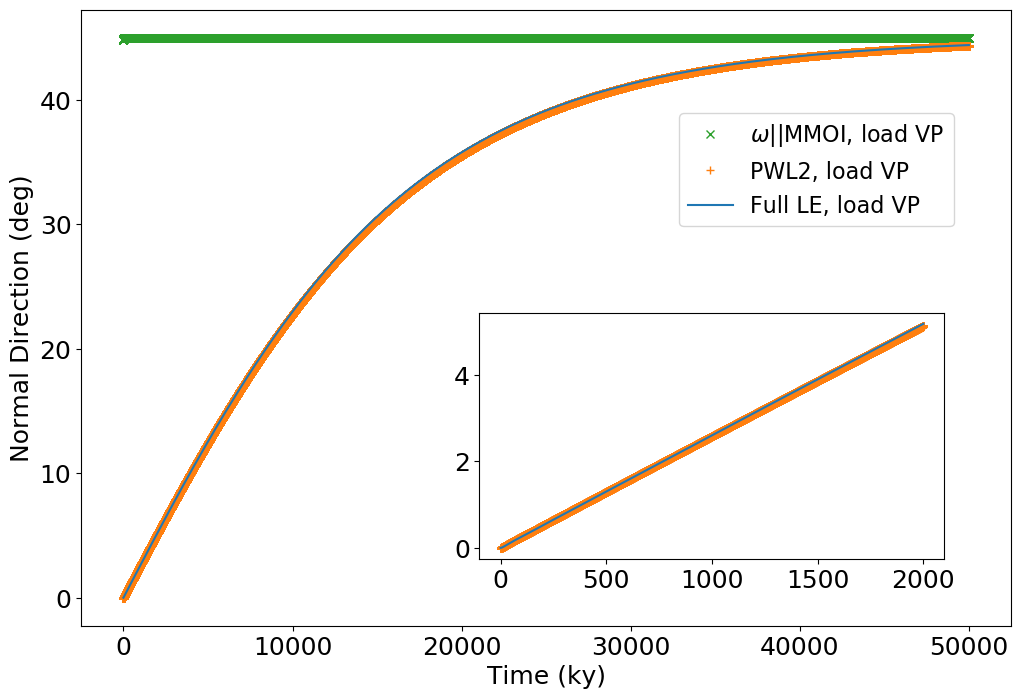}
    \caption{The evolution of TPW in the normal direction for the process studied in Fig.~\ref{figColVP}. The inset of the figure captures the time period studied in Fig.~5 in \citeA{Hu2017b}, and also the load and model parameters are identicals (Tables \ref{tabModels} and \ref{tabLoad}).}
    \label{figNormalDir}
\end{figure}

The PWL2 of \citeA{Hu2017b}, on the other hand, matches the full solution. In Fig.~\ref{figNormalDir} the polar wander is measured in the normal direction, i.e.~the movement of the rotation axis is recorded only in the plane containing the point mass anomaly. In other words, I plot the change of $\bm{n}_\omega \cdot \bm{n}_\mathrm{L}$, with $\bm{n}_\omega$ and $\bm{n}_\mathrm{L}$ denoting unit vectors in the direction of the rotation axis and the load's inertia tensor principal axis respectively. 

Note that I obtain ca.~5 degree motion in the normal direction within the first 2 My, while \citeA{Hu2017b} observe less than 2 degrees (cf.~their Figure 5, top right panel). \citeA{Hu2017b} argue that their solution reveals that the quasi-fluid solution of \citeA{Spada1996} overestimates the normal TPW by more than a factor 4, while it is in fact less than a factor 2 overestimation (\citeA{Spada1996} get ca.~8 degrees within the first 2 My). I suspect that the disagreement between my application of Eq.~23 in \citeA{Hu2017b} and their solution is a result of using overly large time step on their side. \citeA{Hu2017b} claim that a timestep of ca.~5000 years is sufficient to obtain accurate result, but I needed a timestep of ca.~1 year. This is a typical problem when damped oscillations are studied - while the evolution of colatitude may converge relatively soon, a much more fined time stepping is needed to assess the damping correctly (i.e.~the TPW in the normal direction).

\begin{figure}
    \centering
    \includegraphics[width=1.0\textwidth]{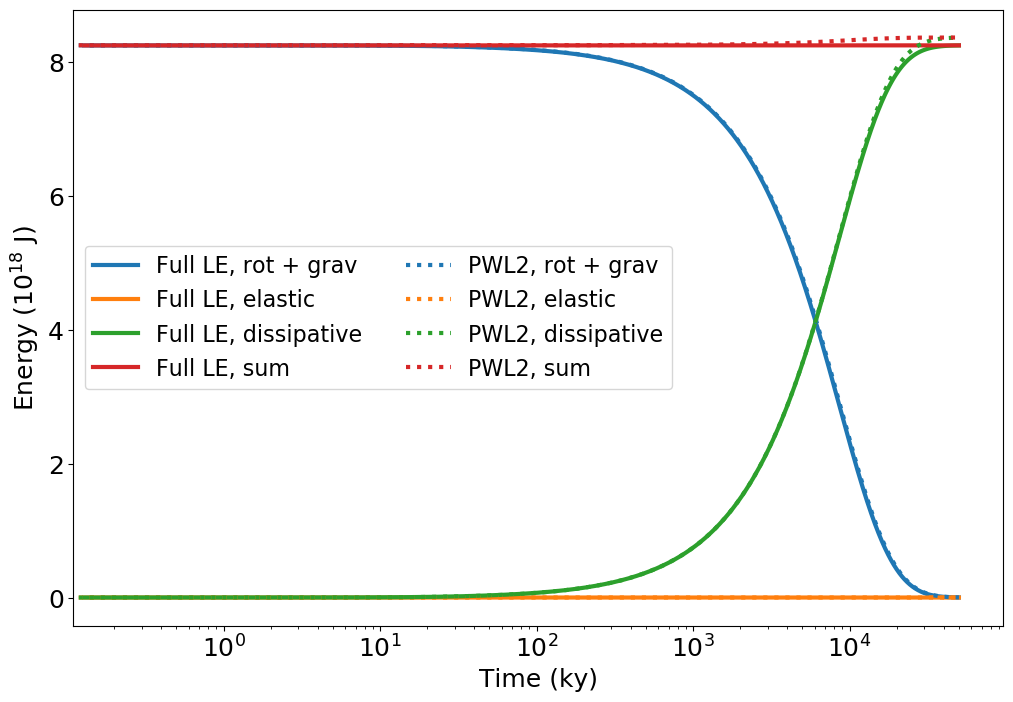}
    \caption{Evolution of the rotational, gravitational, elastic, and dissipative energies for the TPW process studied in Fig.~\ref{figColVP}.}
    \label{figENVP}
\end{figure}

In Fig.~\ref{figENVP} I show the evolution of the underlying energy balance from load emplacement ($t{=}0$) until the wobble is damped and reorientation completed. Again, the energy balance provides a useful check on the solution accuracy, which was employed to confirm the convergence of the full LE solution. The energy is conserved to within \% for the plotted solutions. 

%For fixed loads, the evolution of colatitude resembles a function of the form $C_\mathrm{fin}\,\exp(-t/\tau)$, with $C_\mathrm{fin}$ denoting the final position of $\omg$'s colatitude and $\tau$ denoting a characteristic time scale (cf.~Figs \ref{figColEF} and \ref{figNormalDir}). This may be of use for future studies, if the characteristic time of reorientation is addressed for a range of model parameters.

\section{Discussion}

I have compared the full LE solution with existing approximative methods for two extreme cases: the TPW on a fast rotating planet and the TPW on a slowly rotating planet. For these two endmembers, \citeA{Hu2017a,Hu2017b} develop simplified LE formulations. The main advantage of the full LE approach is, however, no limitation to any endmember scenario in the first place. In this paper I apply LIOUSHELL to previously published test loads, but the code is generally applicable also for cases that lie in between the previously studied endmembers.

The main computational obstacle in predicting long-term TPW is that the full LE intrinsically combines the secular drift with the free oscillations. Numerical integration of the LE is thus a delicate exercise and the energy balance is helpful in providing a better grasp of the achieved accuracy. 

For slowly rotating bodies, the long term TPW and the free oscillations occur on comparable time scales and there is no problem with CPU requirements: e.g.~the 2 My evolution from Fig.~\ref{figColVP} with ca.~$2{\times}10^6$ steps needs less than 30 minutes on a laptop. For a fast rotating body the time scales significantly differ and computing TPW on geological time scales is more challenging, because the numerical time step always has to be only a small fraction of the free wobble period (known as Chandler wobble for Earth).
%Of a particular interest is that for the entire duration of the TPW process there seem to be some free oscillations, although their amplitude is extremely small (cf.~Fig.~\ref{figLongitES}). While this complicates the numerical integration of the full LE, it indicates a link between the planet's long-term TPW and its wobbling. I investigate this link in a follow up paper.

The $\omap$ approximation conserves energy although it entirely omits the wobbling and its damping. In effect, it does not capture the dissipation of the body correctly. This may be a major drawback for slowly rotating planets.

%Worth mentioning is also that energetical analysis quickly reveals inconsistencies such as using values of principal moments of inertia that do not correspond to the hydrostatic shape of the employed model, a problem that has repeatedly been discussed in TPW applications \cite{Hu2017b}

Note that the PWL approach is based on the linearized LE and that the linearized LE strongly violates energy conservation \cite{Patocka2018}. However, the iterative readjustment of coordinates, i.e.~the application of matrix $\bm{Q}$ from Eq.~15 in \citeA{Hu2017a}, seems to remove this deficiency.

A weakness of LIOUSHELL is that tidal forcing is not yet included. This limits its applicability especially for natural satellites with synchronous orbit, as the tidal forcing is even more important than rotation to determine their reorientation. \citeA{Hu2017a} present a method in which the tidal axis can be treated as an additional rotational axis with an ``equivalent angular speed''. Under the assumption that both the tidal and rotation axes always remain perpendicular, they find a solution of the TPW problem with two ``rotational'' axes. In the future I aim to implement the tidal forcing as a term on the right hand side of the LE (cf.~the moment of external forces $\bm{L}$ in Appendix \ref{sec:App}), and to compare the obtained solution with the approach of \citeA{Hu2017a}.

The effect of a fossil bulge is not investigated in this paper. Accounting for a remnant rotational or tidal bulge is, however, a straightforward task within the presented formalism -- it can be added as a fixed load, similarly to the loads EF and VP. In applications, the major problem is in estimating the amplitude and position of such remnant load \cite<see e.g.>{Keane2014}.

\section{Summary}

A general method for modelling TPW on planets is presented. For two endmembers represented by the fast rotating Earth and by the slowly rotating Venus, I compare the solution of the nonlinear LE with solutions obtained using previously published approximative methods.

For fast rotating bodies the long term TPW is accurately captured by the trivial assumption that the maximum principal axis of inertia is always aligned with the rotation axis. The $\omap$ assumption is re-derived and it is related with the contributions of Euler and Coriolis forces to the relative angular momentum of the planet. 

For slowly rotating bodies, the full LE matches the approximative method of \citeA{Hu2017b}. However, the obtained TPW rate in the normal direction seems to be underestimated by more than a factor two in their published solution, as revealed by an analysis of energetics. For planets with neither fast nor slow rotation relative to the planet's free wobble period the full LE must be used.

For fast rotating planets such as Earth, the energy balance of TPW is dominated by a decrease of the rotational energy that is mostly counterbalanced by an increase of the gravitational energy. The elastic energy and dissipation have contributions that are comparable only with the remainder of these two major constituents. For slowly rotating bodies such as Venus the energy balance is dominated by a decrease of the rotational energy that is counterbalanced mostly by the viscous dissipation.

A custom written code LIOUSHELL, used for the calculations in this study, is made available.

\appendix
\section{Role of the Coriolis and Euler forces in derivation of the Liouville equation}\label{sec:App}
Traditionally, the LE is derived by transforming the conservation of angular momentum into the rotating (i.e.~non-inertial) body-fixed reference frame:
\begin{equation}
    \dt{\bm{H}} + \omg \times \bm{H} = \bm{L}
\end{equation}
where $\bm{L}$ is the moment of external forces and the Resal theorem was used to express the time derivative of $\bm{H}$. The theorem holds for any vector $\bm{a}$ that is observed in two mutually rotating frames, $\dt{\bm{a}}|_\mathrm{non-rot} = \dt{\bm{a}}|_\mathrm{rot} + \omg\times\bm{a}$.

However, the balance of angular momentum is valid in all reference frames, and one can thus proceed in a much more tedious, but perhaps also more insightful manner. This alternative path is taken below. 

The balance of angular momentum in the body-fixed frame reads \cite<e.g.>{Munk1960}:
\begin{equation}\label{eqA2}
\dt{\bm{h}} = \int_{v(t)}{\bm{r}\times\bm{f}\,\rho\,\mathrm{d}v ,}
\end{equation}
where $v(t)$ is the volume occupied by the body, $\bm{h}:=\int_{v(t)}{\bm{r}\times\bm{u}\,\rho\,\mathrm{d}v}$ is the relative angular momentum that is measured within the body, $\bm{r}$ is the position vector, and a stress-free surface was assumed. In a non-inertial reference frame, the so-called fictitious forces have to be included as additional body forces $\bm{f}$. Therefore, $\bm{f}$ can be separated into external (e.g.~the gravitational force due to an external body), internal (e.g.~the gravitational force due to mass of the body itself), and fictitious forces: $\bm{f}=\bm{f}_\mathrm{ext}+\bm{f}_\mathrm{int}+\bm{f}_\mathrm{fic}$. After identifying $\int_{v(t)}{\bm{r}\times\bm{f}_\mathrm{ext} \rho \, \mathrm{d}v}$ as the moment of external forces $\bm{L}$, we get:
\begin{equation}\label{eqA3}
\dt{\bm{h}}=\bm{L} + \int_{v(t)}{\bm{r}\times\bm{f}_\mathrm{int} \, \rho \, \mathrm{d}v}+\int_{v(t)}{\bm{r}\times\bm{f}_\mathrm{fic}\, \rho \, dv}
\end{equation}
When all the internal forces are central, the second term on the RHS of Eq.~\eqref{eqA3} is equal to zero. The fictitious forces are \cite<e.g.>{Martinec2019}: 
\begin{equation}
    \bm{f}_\mathrm{fic} = -\omg\times(\omg\times\bm{r})-2\omg\times\bm{v}-\frac{d\omg}{dt}\times\bm{r}-\frac{d\bm{v}_0}{dt}-\omg\times\bm{v}_0
\end{equation}
where $\bm{v}_0$ is the velocity of the origin of the body-fixed frame relative to the inertial reference frame. The last two forces are uniform, and thus their integral contributions to Eq.~\eqref{eqA3} are equal to zero, because in our case the origin of the non-inertial frame lies in the center of mass. Besides $\bm{L}$, the only remaining contributions on the RHS of Eq.~\eqref{eqA3} come from the centrifugal, Coriolis, and Euler forces. Their relative contributions can be written as follows.

Centrifugal force:
\begin{align*}
   -\int_{v(t)}{\bm{r}\times(\omg\times(\omg\times\bm{r}))\,\rho\,\mathrm{d}v} = -\int_{v(t)}{\bm{r}\times(\omg(\omg\cdot\bm{r}) - \bm{r}(\omg\cdot\omg))\,\rho\,\mathrm{d}v} = \\ -\int_{v(t)}{\bm{r}\times \omg(\omg\cdot\bm{r}) \,\rho\,\mathrm{d}v} = \omg\times\int_{v(t)}{\bm{r}\,(\omg\cdot\bm{r}) \,\rho\,\mathrm{d}v} = -\omg\times (\bm{I}\cdot\omg)
\end{align*}

Coriolis force:
\begin{align*}
    - \int_{v(t)}{\bm{r}\times(2\omg\times\bm{v})\,\rho\,\mathrm{d}v} =
    \int_{v(t)}{(\bm{u}(\bm{r}\cdot 2\omg) - 2\omg(\bm{r}\cdot\bm{u}))\,\rho\,\mathrm{d}v} = \\
    -\int_{v(t)}{(\bm{r}(\omg\cdot\bm{u}) - \bm{u}(\bm{r}\cdot 2\omg))\,\rho\,\mathrm{d}v} - 
    \int_{v(t)}{2\omg(\bm{r}\cdot\bm{u})-\bm{r}(\bm{u}\cdot\omg)-(\bm{u}(\bm{r}\cdot \omg))\,\rho\,\mathrm{d}v} = \\
    - \int_{v(t)}{\omg\times(\bm{r}\times\bm{u})\,\rho\,\mathrm{d}v} - \left( \int_{v(t)}{2(\bm{r}\cdot\bm{u})\,\mathbbm{1}-\bm{r}\otimes\bm{u} - \bm{u}\otimes \bm{r})\,\rho\,\mathrm{d}v} \right)\cdot \omg = \\
    -\omg \times \bm{h} - \dt{\bm{I}}\cdot\omega
\end{align*}

Euler force:
\begin{align*}
    -\int_{v(t)}{\bm{r}\times(\frac{d\omg}{dt}\times\bm{r})\,\rho \,\mathrm{d}v} = -\int_{v(t)}{\dt{\omg}r^2 - \bm{r}(\dt{\omg}\cdot\bm{r})\,\rho \,\mathrm{d}v} = -\bm{I}\cdot\dt{\omg}
\end{align*}

Upon inserting these contributions into Eq.~\eqref{eqA3}, the LE is finally obtained:
\begin{equation}
    \frac{d}{dt}(\bm{I}\cdot\omg+\bm{h}) + \omg\times(\bm{I}\cdot\omg+\bm{h})=\bm{L}
\end{equation}
The one to last equalities in Eqs for the centrifugal, Coriolis, and Euler forces come from the definition of the inertia tensor,
\begin{equation}\label{eqIdef}
    \bm{I}=\int_{v(t)}{((\bm{r}\cdot\bm{r})\, \mathbbm{1}-\bm{r}\otimes\bm{r})\rho\, \mathrm{d}v}
\end{equation}
and from its time derivative \cite<e.g.>{Martinec2019},
\begin{equation}
\dt{\bm{I}} = \int_{v(t)}{\left[2(\bm{r}\cdot\bm{v})\,\mathbbm{1}-\bm{r}\otimes\bm{v}-\bm{v}\otimes\bm{r}\right] \rho \, \mathrm{d}v}
\end{equation}
where the latter can be derived by employing Reynold's transport theorem to Eq.~\eqref{eqIdef}.

From the above derivation of the LE it is immediately clear that it is the Coriolis and Euler forces that combine to yield the time derivative terms that appears in the LE. On the other hand, assuming that the centrifugal force is the only fictitious force in the equation of motion directly yields the $\omap$ approximation,
\begin{equation}
    \omg\times(\bm{I}\cdot\omg) + \dt{\bm{h}} = \bm{L}
\end{equation}
Therefore, the range of validity of the $\omap$ approximation is related to the importance of the Euler and Coriolis forces in shaping the body's inertia tensor. Note that the LE is typically solved in the Tisserand frame, i.e.~in the frame with zero relative angular momentum $\bm{h}$ \cite{Munk1960}.

\acknowledgments
I thank Ond\v{r}ej \v{S}r\'{a}mek for comments that helped improve the manuscript and Ond\v{r}ej \v{C}adek for keeping my interest in this topic. The research was supported by OP RDE project No.~CZ.02.2.69/0.0/0.0/18\_053/0016976 International mobility of research, technical
and administrative staff at the Charles University.

%% ------------------------------------------------------------------------ %%
%% References and Citations

%%%%%%%%%%%%%%%%%%%%%%%%%%%%%%%%%%%%%%%%%%%%%%%
%
% \bibliography{<name of your .bib file>} don't specify the file extension
%
% don't specify bibliographystyle
%%%%%%%%%%%%%%%%%%%%%%%%%%%%%%%%%%%%%%%%%%%%%%%

\bibliography{refs}

\end{document}